\documentclass[10pt,journal,compsoc]{IEEEtran}
\newif\ifpeerreview

\peerreviewfalse

\usepackage[nocompress]{cite}
\usepackage{url}
\usepackage{amsmath,amssymb,graphicx}
\usepackage{dsfont}
\usepackage{newtxmath}
\usepackage{booktabs}
\usepackage{multirow}
\usepackage{xcolor}
\usepackage{subcaption}
\usepackage[switch]{lineno}
\usepackage{svg}
\usepackage{float}
\usepackage{caption}
\usepackage{setspace}
\usepackage{makecell}
\usepackage{array}

\newcommand{\paperID}{XXXX}
\newcommand{\Eexpl}{E_{\mathrm{expl}}}

\title{Telemetry is a Sensor:\\ 
Opportunistic Wavefront Estimation for the \\
James Webb Space Telescope}

\author{Lahav Buzi, Yoav Y. Schechner, Aviad Levis and Jason J. Wang%
\IEEEcompsocitemizethanks{
\IEEEcompsocthanksitem L. Buzi and Y. Y. Schechner are with the Viterbi Faculty of Electrical and Computer Engineering, Technion--Israel Institute of Technology, Haifa, Israel.
\IEEEcompsocthanksitem A. Levis is with the University of Toronto, Toronto, ON, Canada.
\IEEEcompsocthanksitem J. J. Wang is with CIERA and the Department of Physics and Astronomy, Northwestern University, Evanston, IL, USA.
}}

\begin{document}

\IEEEtitleabstractindextext{%
\begin{abstract}
Space telescopes maintain optical alignment through periodic, resource-intensive
wavefront calibration, leaving the optical state unobserved between corrections.
We propose treating onboard engineering telemetry as an opportunistic wavefront
sensor. Specifically, the high-cadence thermal and pointing signals already recorded by the James Webb Space Telescope may suffice to recover spatially resolved optical path difference at nanometer precision, without dedicated measurements.
Our framework uses a two-stage gradient boosting regressor. It predicts optical path difference residuals for each mirror segment in a low-dimensional principal component basis. The model is trained and evaluated using data captured on-orbit during six months.
Based on this limited data, optical path difference  inference has high statistical significance in 13 out of 18 mirror segments. In 6 mirror segments, the model achieves explained variance above 50\%. These results demonstrate feasibility for telemetry-driven wavefront
estimation, as a low-overhead sensing modality. The study suggests a potential
to support continuous monitoring and calibration scheduling for large
segmented observatories.
\end{abstract}

\begin{IEEEkeywords}
Computational imaging, wavefront sensing, virtual sensing, space telescopes,
segmented mirrors, machine learning
\end{IEEEkeywords}
}

% Anonymized header for review
\ifpeerreview
\linenumbers \linenumbersep 15pt\relax
\author{Paper ID \paperID\IEEEcompsocitemizethanks{\IEEEcompsocthanksitem This paper is under review for ICCP 2026 and the PAMI special issue on computational photography. Do not distribute.}}
\markboth{Anonymous ICCP 2026 submission ID \paperID}%
{}
\fi
\maketitle

% ======================================================================
%   1. INTRODUCTION
% ======================================================================
\IEEEraisesectionheading{
  \section{Introduction}\label{sec:introduction}
}

\IEEEPARstart{S}{paceborne} observatories must maintain nanometer-scale optical
alignment throughout their lifetimes. Yet, directly measuring optical aberrations
requires pointing the telescope for dedicated calibration observations every
few days. These observations consume observing time that cannot simultaneously be used for science observations. Between these sparse events, the optical state evolves continuously
under thermal loading and structural relaxation, but remains unobserved. The
telescope produces science data under an uncertain wavefront error.
\begin{figure}[t]
\centering
\includegraphics[width=\linewidth]{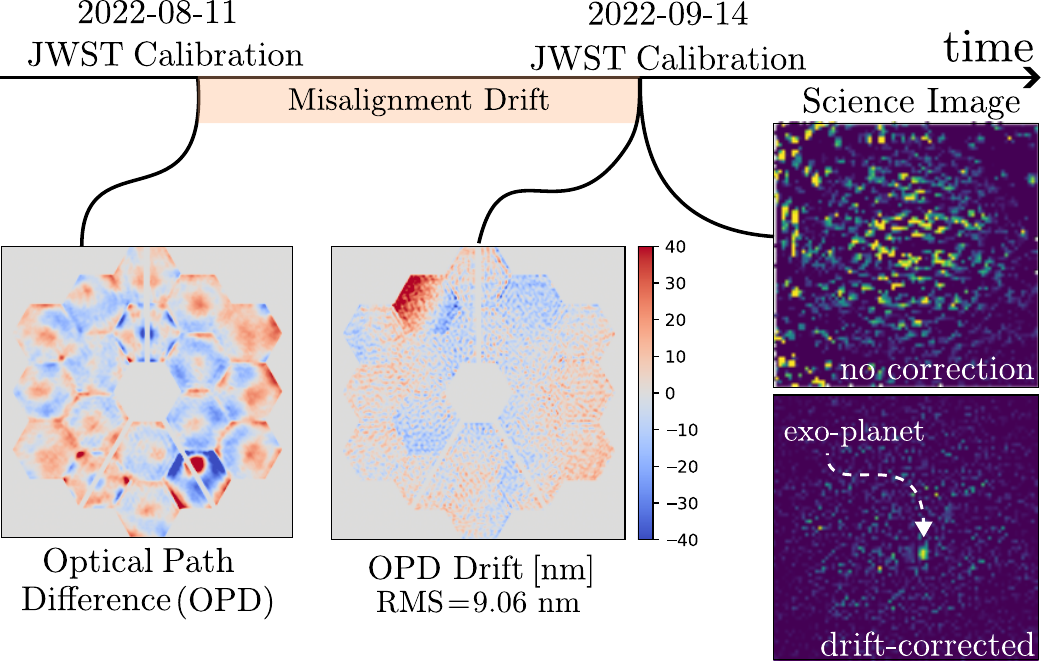}
\caption{
Impact of wavefront drift on high-contrast imaging.
An OPD aberration accumulated between calibration sessions produces speckle artifacts (top-right). This can obscure a faint exoplanet signal. Accurate drift estimation and
removal reveals the companion signal (bottom-right).
}
\label{fig:alignment_star}
\end{figure}
Meanwhile, the optical observatory records high-cadence engineering telemetry. Telemetry includes thermal sensor data, pointing channels, and instrument housekeeping. Some of the engineering telemetry data is affected by physical processes driving optical aberration change. This information, however, is treated as auxiliary diagnostics and is seldom used to infer the optical state.

We investigate this gap by treating telemetry as a \emph{virtual wavefront
sensor}.  Wavefront aberrations are quantified by a two dimensional map of the  optical path difference (OPD).  OPD evolution is governed by thermo-mechanical processes
that telemetry already reflects, at least partially. Hence, we believe, a learned mapping from telemetry to OPD may provide continuous, spatially resolved wavefront estimates without an additional observational overhead. To our knowledge, this work has the first demonstration of recovering spatially resolved, segment-level OPD directly from engineering telemetry, without requiring dedicated wavefront measurements at inference time. We demonstrate this approach on six months of on-orbit data of the James Webb Space Telescope (JWST). This indicates that routine housekeeping signals can support accurate segment-level wavefront reconstruction and science throughput. 
\begin{figure*}[t!]
\centering
\includegraphics[width=0.99\textwidth]{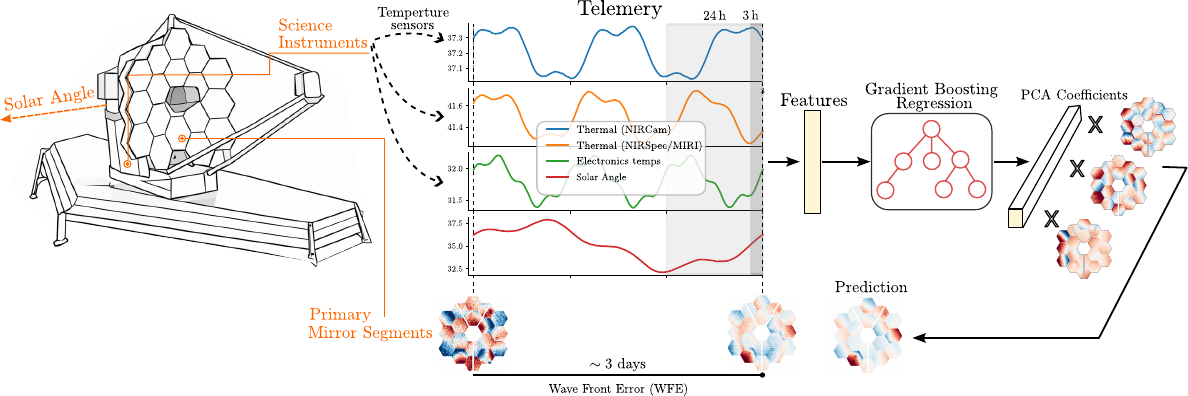}
\caption{
Telemetry-driven wavefront estimation pipeline.
JWST telemetry (temperatures and pointing-related signals) is summarized into features over multiple time windows (e.g., 3\,h and 24\,h) and used to predict low-dimensional PCA coefficients via gradient boosting regression.
These coefficients reconstruct the residual OPD. This approach may enable continuous estimation of the wavefront between sparse sensing measurements.
}
\label{fig:teaser}
\end{figure*}
The imaging motivation is clear. Wavefront drift corrupts high-contrast
observations: a nanometer-scale OPD redistributes energy into point spread function speckles that can obscure faint signals. This is particularly relevant to
exoplanet direct imaging, where drift-induced speckle artifacts reach magnitudes
comparable to a faint companion signal. Improved drift estimation could support
their partial suppression, as illustrated in Fig.~\ref{fig:alignment_star}.

The problem is non-trivial. The JWST has 18 primary-mirror segments. The OPD is routinely calibrated by optical imaging of the mirror. Each mirror segment is observed by  $\sim3\times10^4$ pixels. This provides OPD labeled data.  However, labeled training examples are scarce: we have only 108 calibration measurements over six months. 
Mapping from a set of scalar telemetry readings to high-dimensional wavefront maps is nonlinear and spatially nonuniform, varying between and over segments. This challenge is further compounded by underlying multi-scale temporal dynamics: thermal time
constants in the telescope range from minutes to nearly a day. This may require features in multiple temporal scales. The problem is of high-dimensional regression and virtual sensing, where recovering structured physical states from indirect observations with limited supervision remains challenging~\cite{metzler2017learned,liu2019deep}. 
Our contributions are:
\begin{enumerate}
\item \textbf{Virtual wavefront sensing from telemetry.}
A framework that infers spatially resolved OPD from onboard engineering
telemetry alone, without dedicated wavefront measurements (Fig.~\ref{fig:teaser}).

\item \textbf{Two-stage spatial reconstruction.}
A piston-first, PCA-second pipeline capturing both global mirror-segment drift
and local higher-order deformation across all 18 JWST primary mirror segments.

\item \textbf{On-orbit test.}
Demonstration on six months of JWST data, using a strict time-ordered split
and permutation-based significance testing.
\end{enumerate}

% ======================================================================
%   2. RELATED WORK
% ======================================================================
\section{Related Work}\label{sec:related}

\textbf{JWST wavefront sensing and control.}\quad
JWST's Wavefront Sensing and Control (WFSC) system reconstructs a full-aperture OPD map, via phase retrieval on
defocused stellar images~\cite{acton2012wavefront,perrin2018opd}.
Observatory commissioning involves alignment to $\sim$59\,nm root mean square (RMS). Afterwards, corrections are applied approximately every two days~\cite{feinberg2024stability,rigby2023jwst}.
Drift is primarily thermally driven~\cite{jwstDocsOpticsStability}.
Telfer et al.~\cite{telfer2024empirical} characterize drift statistics
empirically but do not reconstruct the spatial OPD map structure from telemetry.

High-contrast imaging performance is particularly sensitive to residual wavefront errors (aberrations). These errors generate speckle structures that can obscure faint astrophysical signals in adaptive-optics and high-contrast imaging systems~\cite{lewis2023speckle,milli2017adaptive}. Recent computational imaging approaches have further demonstrated the importance of accurate optical and wavefront modeling for exoplanet detection and characterization~\cite{feng2024exoplanet}. Study of these wavefront-induced artifacts provides an additional motivation for improved wavefront monitoring and prediction.

\textbf{Virtual sensing.}\quad
Inferring latent system states from auxiliary measurements is often referred to as virtual or soft sensing. This approach is well studied in monitoring and control~\cite{kadlec2009soft}.
In the astronomical context, telemetry analyses have mainly focused on scalar error
metrics, omitting spatial structure. To our knowledge, no prior work has
demonstrated spatially resolved, segment-level OPD prediction from engineering
telemetry alone. This makes the method a novel application of the virtual sensing paradigm to space telescope optics.

\textbf{Predictive maintenance.}\quad
Operational telemetry has also been used for preventive
maintenance and health monitoring of large observatory
subsystems. Salgado et al.~\cite{salgado2012} show
that long-term telemetry analysis can support maintenance
planning and operational reliability. Their
focus is subsystem maintenance. As in our work, Ref.~\cite{salgado2012} leverages routinely collected engineering telemetry to support operational decision making.

\textbf{Data-driven inverse problems.}\quad
Learning-based methods have shown promise for recovering hidden physical
quantities from indirect or auxiliary observations~\cite{metzler2017learned,
mardani2018deep,liu2019deep}. We apply this approach to optical state
estimation, where the telemetry-to-wavefront mapping is nonlinear
and labeled data are currently scarce.

\textbf{PCA for wavefront representation.}\quad
Principal component analysis (PCA) is used for compact wavefront parameterization~\cite{jolliffe2016pca,
perrin2014updated}. We use per-segment PCA coefficients as regression targets.
This enables learning in a low-dimensional latent space, while preserving spatial
fidelity.

% ======================================================================

%%%%%%%%%%%%%%%%%%%%%%%%%%%5

\section{Data and Telemetry Features}
\label{sec:features}

The JWST primary mirror consists of 18 hexagonal segments arranged
in three concentric rings: inner (A),
middle (B), and outer (C). 
\begin{figure}[t]
\centering
\includegraphics[width=\columnwidth]{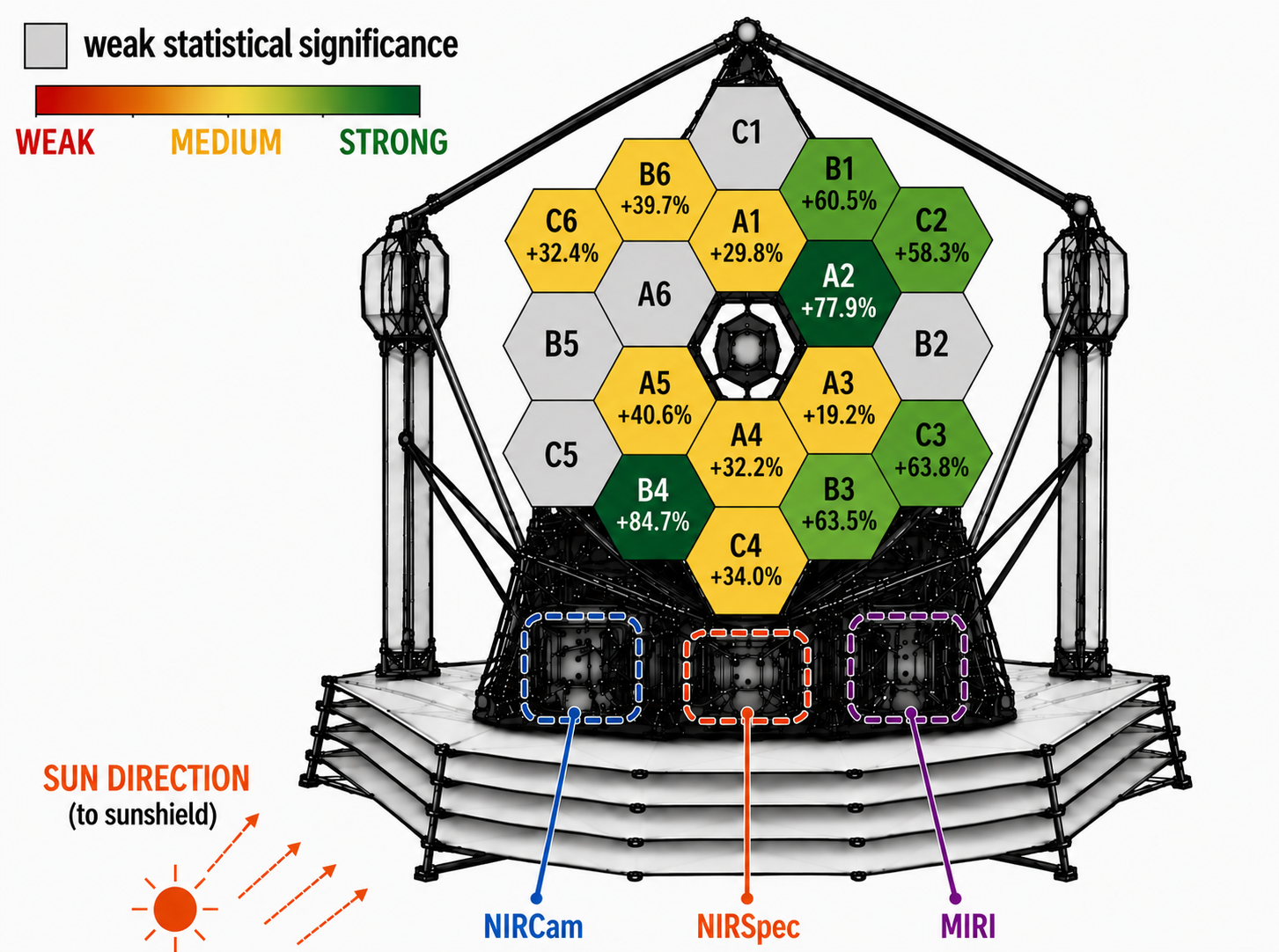}
\caption{
The figure shows the approximate instrument locations and Sun orientation.
The JWST primary mirror has segments labeled (A1-C6).
Most segments denote a corresponding explanation measure   
\(E_{\rm expl}^{\rm obs}\)  (Eq.~\ref{eq:Ematch}) based on a test set exclusive of the train set.
Green segments indicate strong predictive performance,
orange/yellow segments indicate moderate predictive performance.
Gray marks segments where inference does not achieve statistical significance based on the available data.
JWST schematic adapted from publicly available NASA imagery and further modified by the authors.
}
\label{fig:segment_map}
\end{figure}
Fig.~\ref{fig:segment_map} shows  an orientation of the observatory relative to the Sun. JWST is thermally asymmetric due to the sun-shield geometry, and the  telescope orientation affects our interpretation  of segment-level thermo-elastic drift.
For optical labeled data, we have 108 JWST wavefront sensing observations spanning  July-December 2022, acquired at $\sim$2-3 day cadence, reflecting their operational overhead.

\textbf{Telemetry channels.}\quad
The JWST continuously records housekeeping telemetry that reflects its
thermal and operational state.  Fig.~\ref{fig:segment_map}  points  approximate locations of instruments that host some sensors of this state:  the Near-Infrared Camera (NIRCam), Near-Infrared Spectrograph (NIRSpec), and Mid-Infrared Instrument (MIRI). 
The telemetry signals were not designed for OPD sensing (Fig.~\ref{fig:teaser}). We use a subset of the telemetry channels that capture dominant thermal and structural variations~\cite{feinberg2024stability,jwstDocsOpticsStability},
and one fine-guidance count serving as a proxy for Sun angle and telescope pointing.
The channels are listed in Table~\ref{tab:feature_matrix}.
The signal names shown in the table are the official JWST telemetry mnemonics used throughout the mission telemetry archive.

Channel selection was guided by physical intuition. The NIRCam
focal-plane temperature seems highly related to a global OPD drift.
This channel seems to track bulk thermal relaxation of the primary mirror backplane over timescales of hours to days, which was identified in ground
characterization~\cite{feinberg2024stability,jwstDocsOpticsStability} as a significant driver of wavefront changes. NIRSpec and MIRI temperatures seem to relate to 
shorter-timescale segment behavior in adjacent mirror regions. The MIRI telemetry includes measurements from both a cryogenic sensor and a Power Distribution Unit. This relates to internal instrument thermal loading. The fine-guidance count encodes the telescope's Sun angle history. Changes in solar illumination angle alter the heat distribution on the sun-shield and primary mirror assembly. This may drive thermo-elastic deformation on multi-hour timescales. Together, these five channels are affected by  thermal and mechanical degrees of freedom, and are accessible to routine housekeeping telemetry.
%without requiring any channels dedicated to optical sensing.

\begin{table}[t]
\centering
\caption{Telemetry features used as model inputs.
  $T_{\alpha}$ is the lookback window over which statistics are computed.}
\label{tab:feature_matrix}
\footnotesize
\setlength{\tabcolsep}{3pt}
\renewcommand{\arraystretch}{1.1}
\begin{tabular}{
  >{\ttfamily\raggedright\arraybackslash}p{0.34\columnwidth}
  >{\raggedright\arraybackslash}p{0.44\columnwidth}
  >{\centering\arraybackslash}p{0.09\columnwidth}
}
\toprule
\textbf{JWST Signal} & \textbf{Description} & $T_{\alpha}$ \\
\midrule
\makecell[l]{IGDP\_NRC\_FA\_ACE5\\\_SCTEMP}
  & NIRCam focal-plane temperature; dominant thermal driver
  & 24\,h \\
\makecell[l]{IGDP\_NRSI\_C\\\_IFU\_TEMP}
  & NIRSpec integral field unit temperature; instrument thermal state
  & 3\,h \\
\makecell[l]{IMIR\_HK\_ICE\\\_SEN03\_TEMP}
  & MIRI cryogenic sensor; cold-side thermal variation
  & 3\,h \\
\makecell[l]{IMIR\_HK\_PDU\\\_TEMP}
  & MIRI PDU temperature; internal thermal load proxy
  & 3\,h \\
\makecell[l]{SA\_ZFGINSTCT}
  & Fine-guidance count; Sun angle / pointing proxy
  & 3\,h \\
\bottomrule
\end{tabular}
\end{table}

\textbf{Feature construction.}\quad
Formally, let $r_\alpha(t)$ denote the readout of telemetry sensor
$\alpha$ at time $t$. We compute summary statistics over a time window
$T_\alpha$ preceding the prediction time $t$. These include the minimum
$r^{\min}_\alpha$, maximum $r^{\max}_\alpha$, mean
$r^{\mathrm{mean}}_\alpha$, standard deviation
$r^{\mathrm{STD}}_\alpha$, temporal slope
$r^{\mathrm{slope}}_\alpha$, and the last observed value
$r^{\mathrm{last}}_\alpha$.
These statistics form the feature vector used as input to learning
models. Each sensor is associated with either a short-term window
(3\,h) capturing rapid variations or a longer-term window (24\,h)
capturing thermal drift. This reflects the multi-scale dynamics of the system. Overall, the features constitute a 30-dimensional vector $\mathbf g(t)$ at time $t$.
%\begin{equation}
%\mathbf{g}_\alpha(t)=
%\bigl[
%r_\alpha^{\min},
%r_\alpha^{\max},
%r_\alpha^{\rm mean},
%r_\alpha^{\rm std},
%r_\alpha^{\rm slope},
%r_\alpha^{\rm last}
%\bigr]^\top
%\in\mathbb{R}^{6}.
%\end{equation}
%\begin{equation}
%\mathbf g(t)=
%\big[
%\mathbf g_1^\top(t),
%\dots,
%\mathbf g_5^\top(t)
%\big]^\top
%\in\mathbb R^{30}.
%\end{equation}
%The resulting feature vector $\mathbf{g}(t)\in\mathbb{R}^{30}$ constitutes
%the model input at prediction time $t$.
The vector $\mathbf{g}(t)$ is thus not a temporal sequence, but 
a set of statistics in  lookback windows. 

%%%%%%%%%%%%%%%%%%%%%%%%%%%%%%%%%%%%%%%%%%%%%

% ======================================================================
%   4. METHOD
% ======================================================================
%\section{Inferring the OPD}
%\label{sec:method}
% ============================================================================

%\subsection{OPD dimensionality reduction}
\section{OPD Representation}
\label{sec:dim}
\begin{figure*}[t!]
\centering
\includegraphics[width=0.90\textwidth]{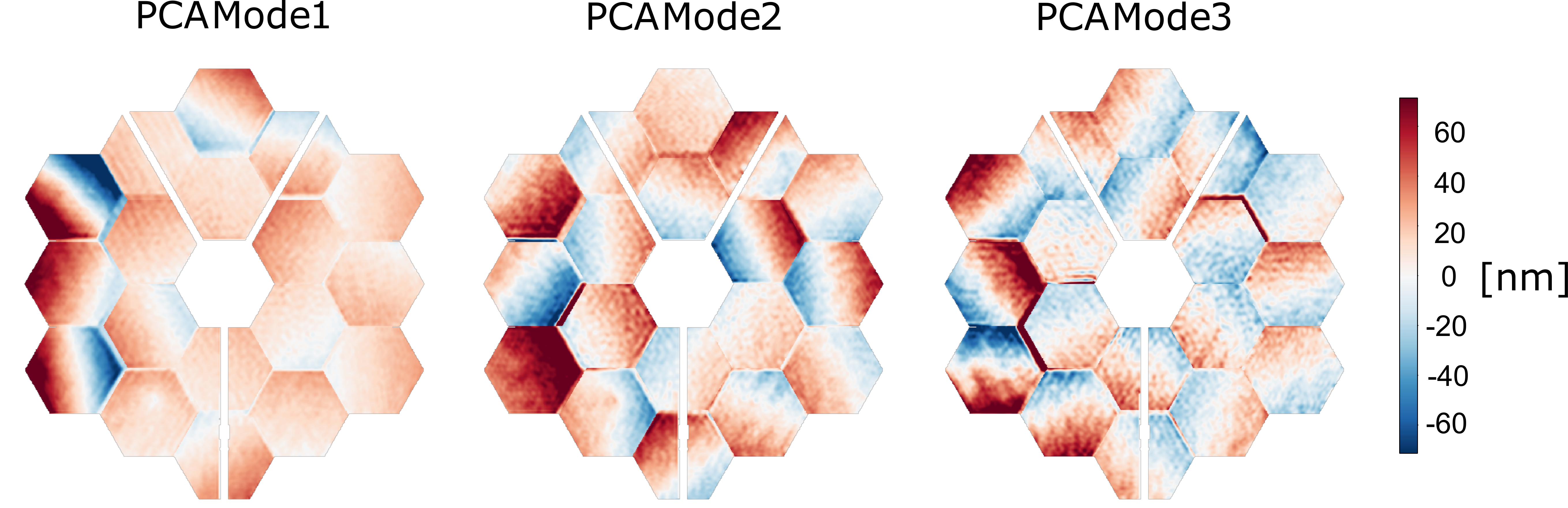}
\caption{%
  Per-segment PCA spatial modes mapped onto the full 18-segment JWST primary
  mirror. %(independent scale per segment).
  Mode~1 is dominated by gradient-like deformation consistent with
  thermal bending; Mode~2 is an orthogonal component; Mode~3 shows
  higher-order curvature. Together, three modes explain $\approx84\%$ of
  the residual variance.}
\label{fig:pca-modes}
\end{figure*}
The OPD at time $t$ is the 2D real-valued map, denoted ${\rm \bf OPD}(t)$.
There is a historical record of measured OPD maps, which label a training set
\begin{equation}
  \Psi=\{{\rm \bf OPD}(t_i)\}_{i=1}^{N^{\rm train}}\;,
\label{eq:Psi}
\end{equation}
where $N^{\rm train}$ is the number of training measurements.
There is also an exclusive test set, whose time samples are denoted ${\xi}\in \Xi$.
There is no temporal overlap between the training and test sets. 
Define the reference OPD as
\begin{equation}
{\rm \bf OPD}^{\rm ref}
=
\frac{1}{N^{\rm train}}
\sum_{i=1}^{N^{\rm train}}
{\rm \bf OPD}(t_i)
\;.
\label{eq:OPDref}
\end{equation}
A residual OPD at time $t_i$ is
\begin{equation}
{\rm \bf OPD}^{\rm residue}(t_i)
=
{\rm \bf OPD}(t_i)
-
{\rm \bf OPD}^{\rm ref}
\;.
\label{eq:OPDresid}
\end{equation}

Over the a mirror spatial domain ${\cal A}$, the RMS of the true residual OPD map is
\begin{equation}
   {\rm RMS}_{\rm OPD}=
   \sqrt{
  \frac{1}{|{\cal A}|}
   \sum_{{\bf x}\in {\cal A}} 
   \left[
     {{\rm  OPD}}^{\rm residue}({\bf x})
   \right]^2
   } \;,
%    \hat{\tilde{\mathbf{a}}}_m(t).
   \label{eq:RMStrue}
\end{equation}
where ${\bf x}$ is a pixel observing the aperture.
Typically, ${\rm  OPD}^{\rm residue}\in[-761,1014]\,{\rm nm}$, with typical per-segment median RMS values
of 9-28\,nm.
In our case, $N^{\rm train}=86$. The JWST OPD map is calibrated occasionally using a megapixel camera, with ${\cal O}(10^5)$ pixels per mirror segment. Without a camera, and with little training data, inferring such a high-dimensional map is a challenge, and typically ill-posed. To handle this, we reduce the dimensionality of the problem. 

Per mirror segment $m$, the corresponding set of pixels of ${\rm \bf OPD}^{\rm residue}(t_i)$ is denoted  ${\rm \bf OPD}^{\rm residue}_m(t_i)$.
The number of pixels in ${\rm \bf OPD}^{\rm residue}_m$ is $N^{\rm pixel}_m$, $\forall i$.
For these pixels, concatenate  ${\rm \bf OPD}_m^{\rm residue}(t_i)$ to a column vector of length  $N^{\rm pixel}_m$, denoted ${\bf o}_{m}(t_i)$. This concatenation is denoted by the operator ${\cal C}$:
\begin{equation}
   {\bf o}_{m}={\cal C}\{ {\rm \bf OPD}_m^{\rm residue}\} 
   \;.
  \label{eq:calC}
\end{equation}
The inverse operator ${\cal C}^{-1}$ rearranges a vector into a 2D map corresponding to mirror segment $m$.

Relative to the telescope frame, each mirror segment $m$ has a spatially-uniform rigid translation along the optical axis. This translation is termed {\em piston}, and denoted $p_m(t)$.  Only the \emph{relative} piston differences between mirror segments affect the point spread function: a global
shift of all mirror segments leaves the wavefront shape unchanged. Inter-segment piston offsets introduce phase discontinuities across the aperture that
broaden the point spread function and reduce the Strehl ratio, a standard measure of optical image quality. 

Let us use one of the mirror segments, denoted $m^{\rm ref}$ as a reference.
Then, the relative piston\footnote{Throughout the paper, a tilde (\(\tilde{\cdot}\)) denotes a quantity after reference subtraction or normalization,
while a hat (\(\hat{\cdot}\)) denotes estimation, e.g., by  model prediction.
For example, \(\tilde{p}_m\) denotes the reference-subtracted
piston, while \(\hat{\tilde{p}}_m\) denotes its estimated or predicted value.} shift is
\begin{equation}
   {\tilde p}_m(t)=p_m(t)- p_{m^{\rm ref}}(t). 
\label{eq:pref}
\end{equation}
For $m=m^{\rm ref}$, ${\tilde p}_m(t)=0$ by definition, $\forall t$.
Estimating and removing piston prevents a dominant uniform offset (per $m$) from monopolizing the subsequent low-dimensional representation of the OPD of $m$. 

%In addition to the piston, each mirror segment has higher-order deformations. This allows spatial modes to represent structured deformation rather than bulk translation.

In addition to the piston, each mirror segment has higher-order deformations. 
%Unlike piston, segment tip and tilt are not spatially uniform offsets.
%They appear as approximately linear gradients across a segment and
%therefore contribute directly to the spatial wavefront structure.
%We consequently retain tip and tilt in the residual OPD representation
%and allow the PCA basis to capture them together with other smooth
%segment deformations. 
The deformations tend to be spatially smooth, and follow particular trends. This is a major key for dimensionality reduction. Each mirror segment has its own trends of deformations. So, dimensionality reduction is done per mirror segment $m$. 
%Representing deformation by a low dimensional basis is part of the training process. 
Various low-dimensional representations of a segment's OPD can be used, including tip/tilt parameters or analytic bases such as Zernike polynomials. We do not require any particular
parameterization. We chose PCA as a data-driven and compact representation that captures the dominant variability
present in the measured OPD residuals. 
%Unlike fixed analytic bases such as Zernike polynomials, PCA is learned directly from measured segment deformations and therefore 
PCA concentrates variance into a small number of coefficients, optimized for the observed JWST residuals. A comparison against Zernike-based parameterization is an interesting direction for future work. 

PCA is a linear representation of the measured OPD residuals.
Photon noise and other image formation effects are inherently nonlinear in the OPD.
However, the present framework operates on calibrated OPD products rather
than on photon count images. Consequently, photon noise enters primarily
as measurement uncertainty in the OPD labels, rather than as an explicit
source of nonlinearity in the mapping of telemetry-to-OPD.

PCA uses the training data $\Psi$ and results in a set of $K\ll N^{\rm pixel}_m$ orthonormal column vectors. Let $\mathbf{1}_m\in\mathbb{R}^{N_m^{\rm pixel}}$ denote an all-ones vector. The PCA vectors are 
\begin{equation}
\{ {\bf v}_{m,k}\}_{k=1}^K=
{\rm PCA}
\left[
\{
{\bf o}_{m}(t_i)
-
{\tilde p}_m(t_i)\mathbf 1_m
\}
\right].
\label{eq:pca}
\end{equation}
These vectors are the $K$ highest principal components. They account for most of the variation of each mirror segment. In our study, $K=3$  modes explain approximately $84\%$ of the residual variance. Additional PCA modes contribute only marginal variance and are therefore omitted to reduce model complexity and mitigate the overfitting risk, given the limited data size. Fig.~\ref{fig:pca-modes} shows the PCA modes. They largely resemble
gradient patterns consistent with tilt, with some higher-order curvature. 

Let $\top$ denote transposition. Define a real-valued vector at any time $t$,
\begin{equation}
   {\bf a}_m(t)
   =
   [a_{m,1}(t),a_{m,2}(t),\ldots,a_{m,K}(t)]^{\top}
   \;.
\label{eq:a}
\end{equation}
At any time $t$, a low-dimensional approximation to the residual OPD of
segment $m$ is
\begin{equation}
{\bf o}_{m}(t)
\approx
{\tilde p}_m(t)\mathbf{1}_m
+
\sum_{k=1}^{K}
a_{m,k}(t)\,{\bf v}_{m,k}
=
{\tilde p}_m(t)\mathbf{1}_m
+
{\bf V}_m{\bf a}_m(t)
\;.
\label{eq:aproxo}
\end{equation}
Here ${\bf V}_m$ is an
$N_m^{\rm pixel}\times K$ matrix,

\begin{equation}
{\bf V}_m
=
[{\bf v}_{m,1},{\bf v}_{m,2},\ldots,{\bf v}_{m,K}]
\;.
\label{eq:V}
\end{equation}

Per mirror segment $m$, the corresponding set of pixels of ${\rm \bf OPD}^{\rm ref}$ is denoted  ${\rm \bf OPD}^{\rm ref}_m$.
Denote an estimate of ${\bf o}_{m}(t)$ by ${\hat {\bf o}}_{m}(t)$. Then, the estimated OPD of mirror segment $m$ is
\begin{equation} 
    \widehat{{\rm \bf OPD}}_m (t) =
    {\rm \bf OPD}^{\rm ref}_m + 
   \widehat{{\rm \bf OPD}}_{m,{\rm residue}} \;,
  \label{eq:estimOPD}
\end{equation}
where 
\begin{equation} 
    \widehat{{\rm \bf OPD}}_{m,{\rm residue}} = 
    {\cal C}^{-1}
      \{  {\hat {\bf o}}_{m}(t) \} \;.
  \label{eq:estimOPDres}
\end{equation}
The full-aperture residual OPD map is obtained by assembling the
18 reconstructed segment maps
\(
\widehat{{\rm \bf OPD}}_{m,{\rm residue}}
\),
yielding
\(
\widehat{{\rm \bf OPD}}_{\rm residue}
\)
over the full aperture ${\cal A}$.
In analogy to Eq.~(\ref{eq:RMStrue}), the RMS of the estimated residual OPD map is
\begin{equation}
   {\rm RMS}_{\widehat {\rm OPD}}=
   \sqrt{
  \frac{1}{|{\cal A}|}
   \sum_{{\bf x}\in {\cal A}} 
   \left[
     \widehat{{\rm  OPD}}_{\rm residue}({\bf x})
   \right]^2
   } \;.
%    \hat{\tilde{\mathbf{a}}}_m(t).
   \label{eq:RMSres}
\end{equation}

From Eq.~(\ref{eq:aproxo}), the OPD estimation problem reduces to estimation of 
${\bf a}_m(t)$ and ${\tilde p}_m(t)$ per mirror segment. 
The number of mirror segments is $N^{\rm seg}$. So, in total, to estimate the OPD map at time $t$, the number of parameters to estimate is 
%$N^{\rm seg}K +(N^{\rm seg}-1)$. 
$N^{\rm seg}(K+1) -1$.
The vector of parameters per $m$ is denoted ${\bf u}_m$. The overall vector of sought parameters is denoted ${\bf u}_{\rm all}$. In our study,  $K=3$ PCA coefficients per segment. Across all $N^{\rm seg}=18$ segments,  the estimation problem reduces from
approximately $5.4\times10^5$ pixel values to only 71 parameters.

%%%%%%%%%%%%%%%%%%%%%%%%%%%%%%%%%%%%%%%%%%%%
\section{OPD Inference}
\label{sec:fit}

To learn the nonlinear telemetry-to-OPD mapping, we use gradient boosting regression (GBR)~\cite{friedman2001greedy}. GBR  is a stage-wise additive model of shallow decision trees. GBR is well suited to small tabular datasets with heterogeneous features and yields interpretable feature importance. Models are trained independently per segment under a strict time-ordered train/test
split to prevent temporal data leakage.

At time $t$, the estimation system receives the feature set appropriate for the prediction task under consideration.
The output is either
$\{ \hat{\bf a}_m(t) \}_m$
or
$\{ \hat{\tilde p}_m(t) \}_m$.
We focus the explanation here on
$\{ \hat{\bf a}_m(t) \}_m$,
for simplicity.

%%%%%%%%%%%%%%%%%%%%%%%%%
\subsection{Gradient-boosting regression}
\label{sec:boost}

\begin{figure}[t]
\centering
\includegraphics[
    width=0.58\columnwidth,
    trim=0.6cm 0.4cm 0.6cm 0.4cm,
    clip
]{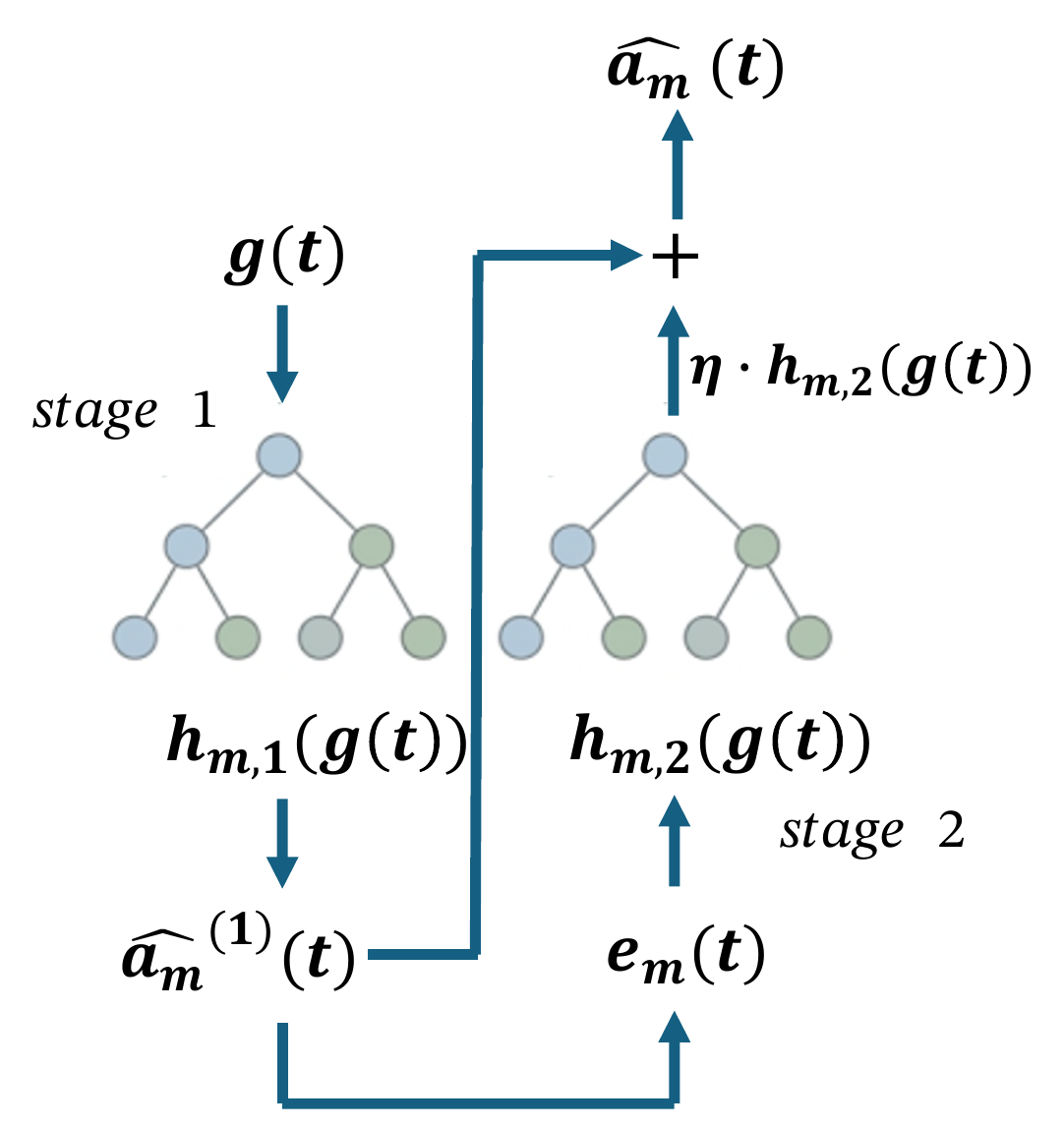}
\caption{%
Illustration of gradient boosting regression (GBR).
Each boosting iteration adds a shallow regression tree that predicts
a correction to the current estimate in PCA-coefficient space.
The final prediction is obtained as the sum of all boosting iterations.
This yields a nonlinear mapping from telemetry inputs to
$\hat{\mathbf{a}}_m(t)$.}
\label{fig:gbr}
\end{figure}

For any $t$, let the time invariant function $h_{m,1}$ be a simple estimator of ${\bf a}_m(t)$, i.e.,
\begin{equation}
    {\bf a}_m(t) \approx h_{m,1} [{\bf g}(t); {\Phi}_{m,1}]
    \;.
  \label{eq:h1m}
\end{equation}
Here ${\Phi}_{m,1}$ is a set of parameters that control this function. The estimation error is
\begin{equation}
    {\bf e}_m(t)= 
    {\bf a}_m(t) -h_{m,1} [{\bf g}(t); {\Phi}_{m,1}]
    \;.
  \label{eq:e1m}
\end{equation}
Then, for $l>1$ let the time invariant function $h_{m,l}$ be a simple estimator of ${\bf e}_m(t)$, i.e.,
\begin{equation}
    {\bf e}_m(t) \approx h_{m,l} [{\bf g}(t); {\Phi}_{m,l}]
    \;.
  \label{eq:elm}
\end{equation}
Here ${\Phi}_{m,l}$ is a set of parameters that control the function $h_{m,l}$. After iteratively defining and using $L$ such simple functions, the estimation is
\begin{equation}
    %{\bf e}_m(t)=  
    \hat {\bf a}_m(t) \approx
    h_{m,1} [{\bf g}(t); {\Phi}_{m,1}] +
    \eta
    \sum_{l=2}^{L}
    h_{m,l} [{\bf g}(t); {\Phi}_{m,l}]
    \;.
  \label{eq:umL}
\end{equation}
Here $\eta$ is a {\em learning rate}. Overall, the model infers an accumulated sum of small corrections, each contributed by a simple parametric function. The boosting process is illustrated in Fig.~\ref{fig:gbr}. Here, each boosting iteration corresponds to a shallow regression tree within the GBR ensemble.

%%%%%%%%%%%%%%%%%%%%%%%%%
\subsection{Decision Tree}
\label{sec:tree}

\begin{figure*}[t!]
\centering
\includegraphics[width=0.92\textwidth]{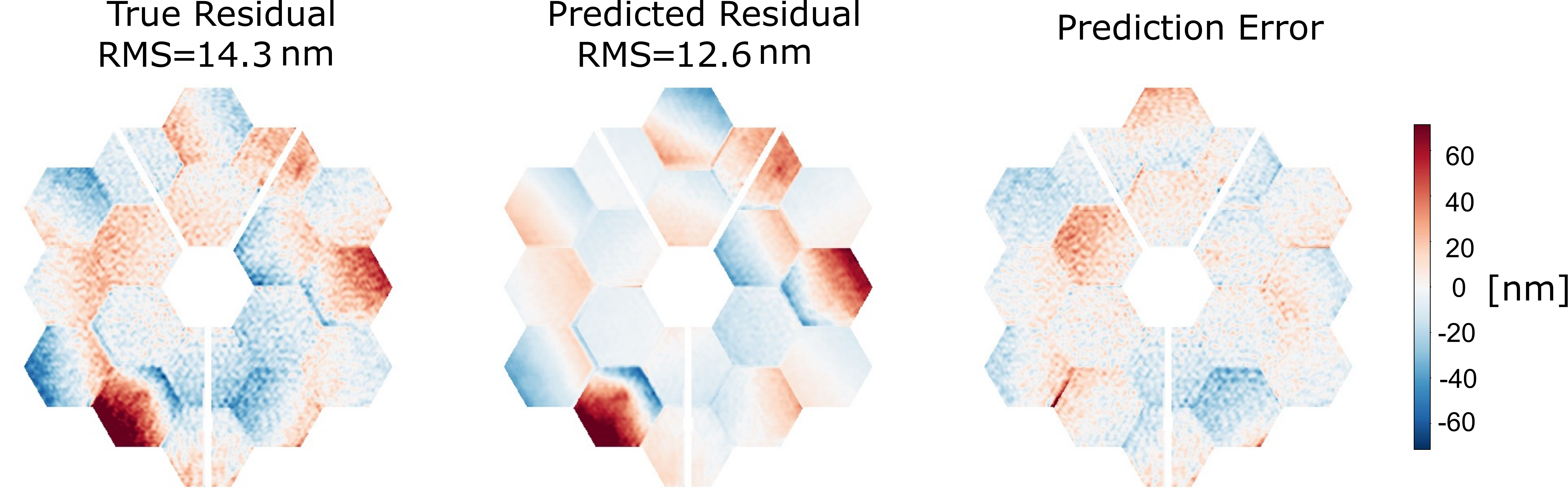}
\caption{%
  An example of a reconstruction of the full-aperture residual OPD 
  residual $ \widehat{{\rm \bf OPD}}_{\rm residue}$.
  The model captures the dominant large-scale wavefront structure across all  18 segments.}
\label{fig:good-day}
\end{figure*}
We opted that each simple function in the set $\{ h_{m,l} \}$ would be a decision tree. The tree has a sequence of splits. Index a split in a tree by $s$.
Each split decides which single feature to use, say element $q_s\in [1,\ldots,N^{\rm feature}]$ of feature vector ${\bf g}(t)$, denoted
$g_{q_{m,l,s}}(t)$. Then, a threshold $\theta_{m,l,s}$ is set. 

Suppose $g_{q_{m,l,s}}(t)<\theta_{m,l,s}$. Then, there are two possible outcomes.
One possible outcome is an actual output vector. Generally, this is an estimated error vector $\hat{\bf e}_{m,l,s}^{<}$ for a tree $h_{m,l}$, where $l>1$. 
If $l=1$, an output vector of the tree $h_{m,1}$ is  $\hat{\bf a}_{m,l,s}^{<}$.
Alternatively, the outcome is a new split, indexed $s+1$. A new split is performed, if there is another feature $q'$ and/or another threshold
$\theta_{m,l,s+1}$, which improves approximation of the output vector.
Similarly, if  $g_{q_{m,l,s}}(t)\geq\theta_{m,l,s}$, the result is either an estimated output vector $\hat{\bf e}_{m,l,s}^{>}$ or a new split.

The set of parameters of tree $h_{m,l}$ is thus 
\begin{equation}
 \Phi_{m,l}=\{ q_{m,l,s}, \theta_{m,l,s} ,\hat{\bf e}_{m,l,s}^{<}, \hat{\bf e}_{m,l,s}^{>} \}_s
    \;.
  \label{eq:phiml}
\end{equation}
This set is determined by training. Training is done by the optimization
\begin{equation}
\hat\Phi_{m,l}
=
\arg\min_{\Phi_{m,l}}
\sum_{i=1}^{N^{\rm train}}
\|
{\bf e}_m(t_i)
-
h_{m,l}[{\bf g}(t_i);\Phi_{m,l}]
\|^2
\;,
\label{eq:trainphi}
\end{equation}
over the training set. 
We use a standard off-the-shelf implementation of Gradient Boosting Regression as provided in the \texttt{scikit-learn} library~\cite{pedregosa2011sklearn}. 
%~\cite{pedregosa2011scikit}.
The model consists of an ensemble of shallow regression trees trained in a stage-wise manner, following the formulation of Friedman~\cite{friedman2001greedy}. We use shallow trees (depth $\approx 3$) and a small learning rate, for stable stage-wise refinement.

%%%%%%%%%%%%%%%%%%%%%%%%%%%%%%%%%%%%%%%%%%%%
\subsection{Estimation in two stages}
\label{sec:2stage}

Piston is a major contributor to the OPD.
For this reason, we first estimate
$\{ \tilde{p}_m(t) \}_m$
using a GBR model for each segment $m$.
The input consists of telemetry features together with the most recently measured piston value for that segment, obtained at an earlier time $t'<t$.
The resulting estimate per model (and mirror segment) $m$ is denoted
$\hat{\tilde p}_m(t)$.

We then estimate
$\{ {\bf a}_m(t) \}_m$
using a GBR model for each segment $m$.
The input consists of the above mentioned telemetry features, together with the complete set of estimated relative pistons
$\{ \hat{\tilde p}_m(t) \}_m$
obtained in the first stage. 

\begin{figure}[t]
\centering
\includegraphics[width=0.95\columnwidth]{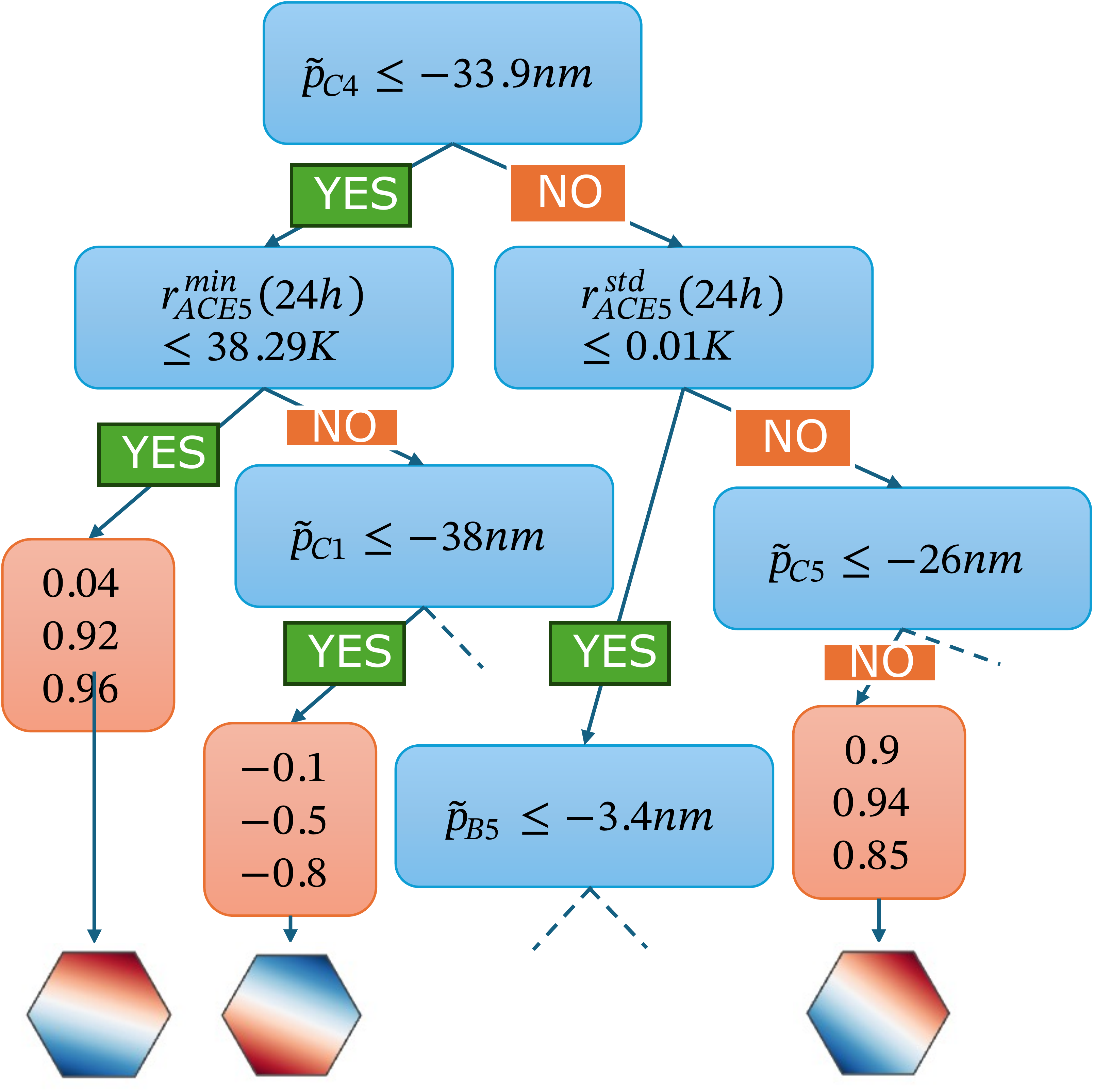}
\caption{%
  A regression tree from the GBR sequence for mirror segment C6.
  Each node applies a threshold on a telemetry or inter-segment
  piston feature, partitioning the input feature space into distinct regimes. A leaf outputs a vector of  PCA coefficients,
  inferring the OPD for C6.}
\label{fig:tree-C6}
\end{figure}

Figure~\ref{fig:good-day} shows an example of the inferred residual OPD map,  $\widehat{{\rm \bf OPD}}_{\rm residue}$, in a test. 
An example tree extracted from the GBR sequence
is shown in Fig.~\ref{fig:tree-C6}. Each internal node splits on a telemetry or piston feature. Eventually, each leaf outputs the $K=3$ PCA coefficients $[\hat{a}_{m,1},\hat{a}_{m,2},\hat{a}_{m,3}]$ simultaneously, leading to the OPD map. 

Our main tools to counter overfitting are dimensionality 
reduction and simple models.
In the first stage, piston estimation, there are 17 simple models: one per mirror segment $m$ (excluding the reference mirror segment). 
In each of these models, there is only a single value to to infer:
$\tilde{p}_m(t)$.
For this, the model trains on $N^{\rm train}=86$ labeled time samples $i$, each having a 30-dimensional feature vector ${\bf g}(t_i)$ and a ground truth relative piston value. In addition, the model has a datum of the recent measured piston state. 

In the second stage, estimation of segment deformation, there are distinct, independent models: one per mirror segment. Per segment, the model
infers just 3 unknowns (PCA coefficients). For this, the model trains on $N^{\rm train}=86$ labeled time samples $i$, each having a 30-dimensional feature vector ${\bf g}(t_i)$ and 3 ground truth PCA coefficients on these time samples. In addition, the model has as input 17 values of the estimated piston state, across the mirror. 

We assess that for each of these models, there is a favorable balance of unknowns to infer  vs.~the data size. This is one reason that we did not apply cross-validation within the training set. Moreover, 
within our limited dataset, random cross-validation risks mixing measurements between the training and test sets, leading to information leakage. This risk is difficult to mitigate with so little data. Hence, we opted for a time-ordered split. Training by cross validation can, however, be carefully tried here, and this may potentially improve performance. 

It is possible to estimate the residual OPD in a single stage. In this approach, both the piston and the PCA coefficients of mirror segment $m$ are estimated simultaneously. For each mirror segment $m$, the model outputs a four-element vector. The results are similar to those obtained with the two-stage approach. Nevertheless, we believe the two-stage approach is beneficial because it provides more interpretable results.

% --------------------------
\section{Interpretable Trees}
\label{sec:surro}
% --------------------------

Nonlinear inference models are often difficult to interpret and gain insights from. There is a way to obtain an interpretable model in our case: a surrogate tree. The surrogate tree has the following characteristics: (a) It is a single tree per $m$, not using the GBR algorithm. 
(b) The tree for segment $m$ does not train on the true OPD or the corresponding ${\bf a}_m(t)$. Instead, it trains on the vectors $\hat {\bf a}_m(t)$, which  are the final output inferred by Eq.~(\ref{eq:umL}). (c) The tree is shallow, to help human interpretation. In our case, it creates a set of several simple threshold-based rules, in telemetry space. We use a surrogate tree having a depth of three decision levels. 

The surrogate tree is a nonlinear function $h^{\rm sur}_m$. It 
results in an approximated inference, 
\begin{equation}
   \hat{\mathbf{a}}^{\rm sur}_m(t)=
  h^{\rm sur}_m[\mathbf{g}(t)]
    \approx
     \hat{\mathbf{a}}_m(t).
%    \hat{\tilde{\mathbf{a}}}_m(t).
   \label{eq:sura}
\end{equation}
Each leaf maps a telemetry regime, defined by threshold conditions on individual channels, to a predicted PCA coefficient vector. In our experiments, the surrogate tree explains approximately 65\% of the variance of the original GBR predictions while remaining human-readable. The surrogate model is not used for OPD prediction itself, but only as an interpretable approximation of the trained GBR model, providing a simplified view of the dominant decision rules learned by the full model.

\begin{figure}[t]
\centering
\includegraphics[width=0.95\columnwidth]{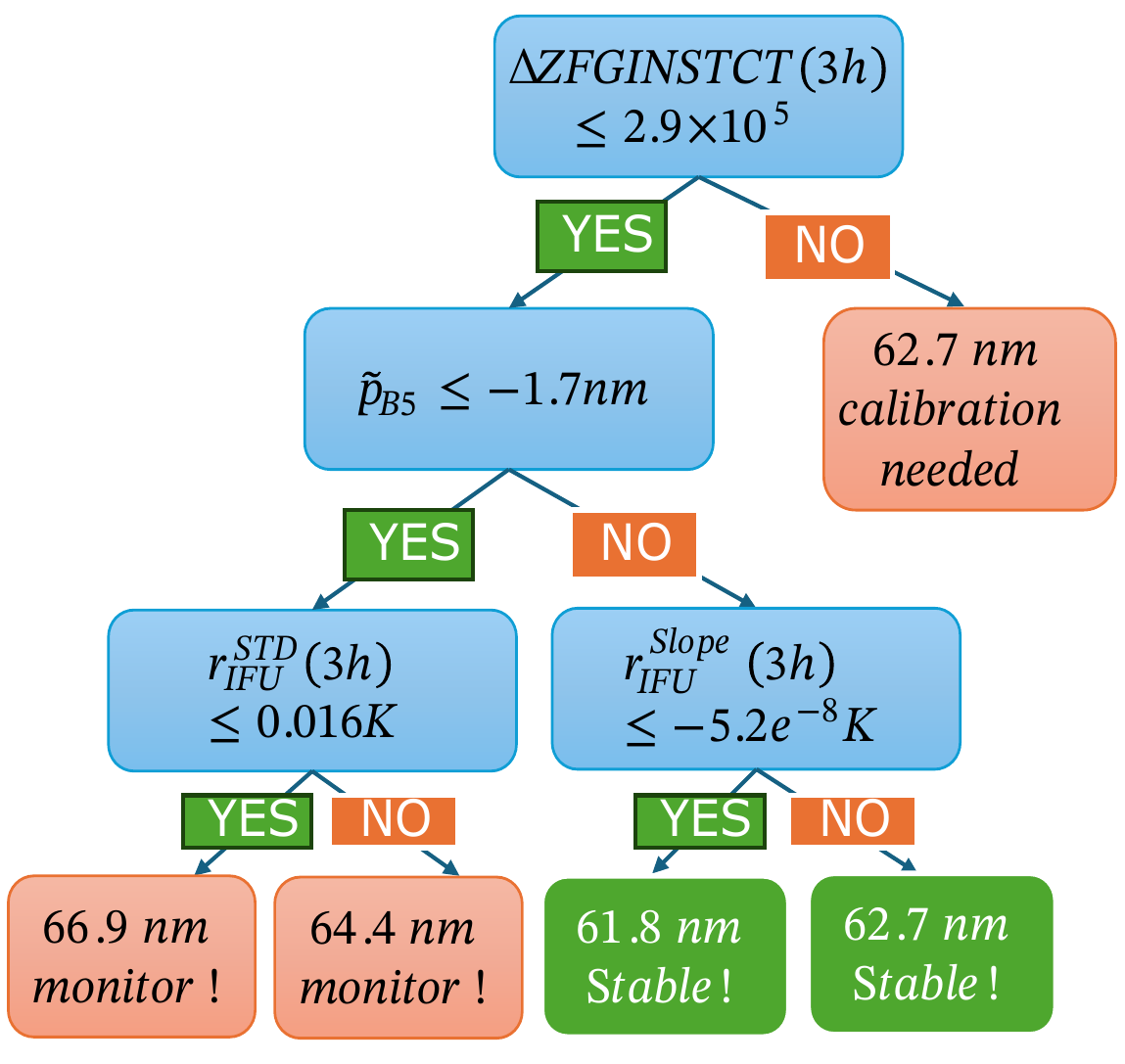}
\caption{%
  A surrogate decision tree for segment B6.
  The tree leaves yield estimated residual OPD maps. 
 Each map yields a predicted ${\rm RMS}_{\widehat {\rm OPD}}$ in nm. This value indicates a course of action for the observatory operators.}
\label{fig:tree_b6}
\end{figure}

An additional product is to deliver a decision tree, which instructs the JWST regarding an action, such as the need to calibrate the OPD.
Specifically, current operation requires re-alignment of the mirror segments, if calibration shows that 
${\rm RMS}_{{\rm OPD}}\geq 60$nm at time $t$. However, often, an OPD calibration session comes at the expense of scientific observations, sometimes  finding out that the mirror is still in a reasonable state. 
The approach of this paper can make the process more efficient. Suppose that inference indicates that at time $t$, ${\rm RMS}_{\widehat {\rm OPD}}$ is much lower. Then, there is likely no need to stop scientific observations for a calibration session. An example focusing on a single mirror segment is shown in Fig.~\ref{fig:tree_b6}. 
This action-oriented tree is based on a surrogate decision tree for approximating the inferred OPD.

%%%%%%%%%%%%%%%%%%%%%%
\section{Statistical Significance}
\label{sec:Signi}

Let $\Xi$ be the time instances of a test set. 
A test sample is taken at time 
${\xi}\in \Xi$.  The estimation mean absolute error (MAE) is
\begin{equation}
{\rm MAE}=
\frac{1}{|\Xi|\,|{\cal A}|}
\sum_{\xi\in\Xi}
\sum_{{\bf x}\in{\cal A}}
\left|
{\rm OPD}^{\rm residue}({\bf x},\xi)
-
\widehat{{\rm OPD}}_{\rm residue}({\bf x},\xi)
\right|.
\label{eq:MAE}
\end{equation}
The estimation error RMS is 
\begin{equation}
{\rm RMS}_{\rm E}(\xi)
=
\sqrt{
\frac{1}{|{\cal A}|}
\sum_{{\bf x}\in{\cal A}}
\left[
{\rm OPD}^{\rm residue}({\bf x},\xi)
-
\widehat{{\rm OPD}}_{\rm residue}({\bf x},\xi)
\right]^2
}
.
\label{eq:Error}
\end{equation}
Based on Eqs.~(\ref{eq:RMStrue},\ref{eq:Error}), define real-valued
{\em explanation} criteria
\begin{equation}
  \Eexpl(\xi)=
  1-\!\left[
  \frac{{\rm RMS}_{\rm E}(\xi)}
       {{\rm RMS}_{\rm OPD}(\xi)}
  \right]^{\!2},
    \label{eq:Ematchxhi}
\end{equation}
\begin{equation}
  \Eexpl= \frac{1}{|\Xi|}
   \sum_{\xi\in \Xi} \Eexpl(\xi).
    \label{eq:Ematch}
\end{equation}
The special case $\Eexpl=1$ corresponds to perfect reconstruction.
On the other, a special case $\Eexpl=0$ corresponds to errors as large as the sought map. A negative $\Eexpl$ means that the errors surpass the sought values. For inference based on temporally-corresponding data, as we derive in this paper,  the result of Eq.~(\ref{eq:Ematch}) is denoted $\Eexpl^{\mathrm{obs}}$.

Suppose $\Eexpl^{\mathrm{obs}}=0.84$. Does this indicate a good fit, statistically? 
The number of measured OPD maps available to us is limited. So, we assess statistical significance using permutation tests. Permutation testing is a non-parametric approach for assessing whether an observed prediction score is significantly better than would be expected under a null hypothesis of no relationship between inputs and targets~\cite{good2005permutation}.
A null hypothesis is that there is no
relation between telemetry features (input) ${\bf g}(t_i)$ and the sought OPD representation vector $\mathbf{a}_m(t_i)$ of a mirror segment $m$ (target). Permutations assess this hypothesis. 

Let us randomly permute time samples, only in the {\em training set} $\Psi$. In a permutation indexed $j$, each time sample $i$ is replaced by a random time sample denoted $i'(j)$, where both are from the training set.  Now, let us {\em re-train} a GBR model, to fit  $\mathbf{a}_m(t_{i'(j)})$ to ${\bf g}(t_i)$, using the process described in Sec.~\ref{sec:fit}. 
Then, the re-trained model infers $\widehat{\bf OPD}_{\rm residue}(\xi)$ in the test set ${\xi}\in \Xi$. Applying Eqs.~(\ref{eq:Error},\ref{eq:Ematchxhi},\ref{eq:Ematch}) on this result yields an explanation measure denoted 
$\Eexpl^{(j)}$.

There are $N^{\rm permute}$ permutations. 
Let $\mathds{1}$ be the indicator function, having value 1 iff its argument is true.
Then, the empirical {\tt p}-value is:
\begin{equation}
   {\tt p}_{\mathrm{rand}} =
     \frac{1}{N^{\rm permute}}
    \sum_{j=1}^{N^{\rm permute}}
    \mathds{1}  
     \left\{
      \Eexpl^{(j)} \geq \Eexpl^{\mathrm{obs}}
     \right\}\;.
   \label{eq:prand}  
\end{equation}
For a statistically significant estimator, the chance is very small that a random permutation in time would improve the explanation criterion, hence  
${\tt p}_{\mathrm{rand}}$ is small. 

%We define the explanation metric on a per-sample basis rather than as a single global statistic in order to evaluate predictive reliability at individual operational time instances. This choice assigns equal weight to each test observation, independent of its absolute OPD magnitude. None of the evaluated on-orbit test samples exhibited near-zero ${\rm RMS}_{\rm OPD}$ values (typical segment median values range from 9 to 28\,nm), and the per-sample normalization therefore remained numerically stable throughout the evaluation. 

To account for temporal correlations, we additionally perform a
block permutation test, where contiguous time windows are shuffled. In analogy to 
Eq.~(\ref{eq:prand}), block-permutations yield a measure ${\tt p}_{\mathrm{block}}$.
Following standard permutation-testing practice~\cite{good2005permutation}, we consider our inference to be 
statistically significant if ${\tt p}_{\mathrm{rand}}<0.05$ and 
${\tt p}_{\mathrm{block}}<0.1$. 

The values of MAE and $\Eexpl^{\mathrm{obs}}$ per mirror segment are listed in 
Table~\ref{tab:time-ordered}. The $\Eexpl^{\mathrm{obs}}$ values also appear in
Fig.~\ref{fig:segment_map}.
\begin{table}[t]
\renewcommand{\arraystretch}{1.15}
\caption{%
  Performance statistics of our model, evaluated on a test set exclusive of the train set. Horizontal rules separate highly statistically significant (top), weakly significant (middle), and not statistically significant (bottom) segment subsets. Segment labels (A1-C6) correspond to the JWST primary-mirror segments shown in Fig.~\ref{fig:segment_map}.}
\label{tab:time-ordered}
\centering
\small
\begin{tabular}{l r r}
\toprule
Seg & $\Eexpl^{\mathrm{obs}}$ [\%] & MAE [nm] \\
\midrule
B4 & 84.7 &  7.1 \\
A2 & 77.9 & 10.5 \\
C3 & 63.8 & 10.1 \\
B3 & 63.5 & 11.1 \\
B1 & 60.5 &  8.4 \\
C2 & 58.3 &  5.5 \\
A5 & 40.6 &  7.5 \\
B6 & 39.7 &  7.7 \\
C4 & 34.0 &  7.0 \\
C6 & 32.4 &  8.3 \\
A4 & 32.2 & 15.4 \\
A1 & 29.8 &  9.3 \\
A3 & 19.2 &  5.8 \\
\midrule
C5 & 10.4 & 16.5 \\
\midrule
B2 &  0.1 & 12.7 \\
A6 & -3.6 &  6.3 \\
C1 & -3.8 &  8.0 \\
B5 & -18.9 &  7.0 \\
\bottomrule
\end{tabular}
\end{table}
Four segments (A6, B2, C1 and B5) do not achieve statistical significance under the adopted permutation criteria.

% ======================================================================
%   5. RESULTS
% ======================================================================

%\begin{figure}[t]
%\centering
%\includegraphics[width=\columnwidth]{Fig5_resid_vs_error_scatter_bold.pdf}
%\caption{%
%  Residual RMS vs.\ prediction error RMS per test sample.
%  Points below the diagonal (dashed) indicate improvement over the
%  mean-only baseline.
%  Validated segments (green) show consistent improvement, with the largest
%  gains in high-residual-RMS samples.
%  Non-validated segments (gray) do not improve reliably.}
%\label{fig:scatter}
%\end{figure}

%\begin{figure}[t]
%\centering
%\includegraphics[width=\columnwidth]{Fig6_map_examples_strong_vs_nosignal.pdf}
%\caption{%
%  Representative segment reconstructions (residual / predicted / error maps).
%  \emph{Top}: a segment with strong spatial gradient structure;
%  $\Eexpl\approx94\%$, error $\approx8$\,nm.
%  \emph{Bottom}: a near-uniform, piston-dominated segment where telemetry
%  carries little spatial information and the model does not improve over
%  baseline. The contrast highlights the fundamental observability condition:
%  spatial structure is required for telemetry-driven prediction.}
%\label{fig:good-bad}
%\end{figure}

%%%%%%%%%%%%%%%%%%%%%%%%%%%%%%%%%%%%%%%555

\section{Other Regression Models}
\label{sec:model_comparison}

GBR, of course, is not the only possible algorithm for model fitting.
We did not seek deep neural architectures, due to 
the limited number of labeled OPD measurements currently 
available to us. 
We tried several other, widely used regression models under identical
 settings, time-ordered train/test split and PCA bases. 
The random forest (RF)~\cite{breiman2001random} uses an ensemble of
independently trained decision trees. This model  has comparable motives to 
GBR. Support vector regression (SVR)~\cite{drucker1997support} has a nonlinear kernel-based model. It is useful for small datasets. K-nearest neighbors
(KNN)~\cite{cover1967nearest} is a non-parametric local fitting function, relying on similarity in feature space. Ridge~\cite{hoerl1970ridge} and Lasso~\cite{tibshirani1996lasso} regressions use linear models, with
$\ell_2$ and $\ell_1$ regularization, respectively.  
Histogram Gradient Boosting Regression (HGBR)~\cite{pedregosa2011sklearn}
is a modern boosting-based approach. 

This comparison has model fitting that removes $p_{m^{\rm ref}}$, and does not exploit the pistons in the estimation of the OPD PCA coefficients.  Hence, all comparisons rely solely on telemetry here. The results are summarized in Table~\ref{tab:model_comparison}.
\begin{table}[t]
\centering
\caption{
Comparison of regression methods.
The $3^{\rm rd}$ column reports the number of segments having
$E_{\rm expl}^{\rm obs}>50\%$.
}
\label{tab:model_comparison}
\small
\begin{tabular}{lccc}
\toprule
Model & $\Eexpl^{\rm obs}$ [\%] & Segments $>50\%$ & MAE [nm] \\
\midrule
GBR   & \textbf{33.5} & \textbf{6} & \textbf{12.4} \\
RF    & 32.2 & 4 & 13.2 \\
SVR   & 14.9 & 1 & 14.8 \\
KNN   & 12.6 & 2 & 14.8 \\
Lasso & 6.0 & 2 & 15.3 \\
Ridge & 3.6 & 2 & 15.4 \\
HGBR  & 0.2 & 2 & 15.9 \\
\bottomrule
\end{tabular}
\end{table}
The table indicates that telemetry-to-OPD regression is
nonlinear and benefits from GBR.

%%%%%%%%%%%%%%%%%%%%%%%%%%%%%%%%%%%%%%%%%%%%%%%

\section{Robustness of GBR to Perturbations}
\label{sec:robustness}

GBR is composed of threshold-based decisions. So, we evaluate whether crossing a learned split threshold can introduce major discontinuities in OPD predictions. 
We perturb a feature across a learned GBR threshold, while holding
all remaining input features fixed. Specifically, a selected feature, denoted
$g$ is shifted from slightly below to slightly above the threshold in a tree split indexed $s$:
\begin{equation}
g^-=\theta-\epsilon,
\qquad
g^+=\theta+\epsilon.
\label{eq:threshold_crossing}
\end{equation}
Here \(\theta\) denotes the learned split threshold and
\(\epsilon\) is a small perturbation magnitude. All remaining components of \(\mathbf g\) are held fixed.
For telemetry-derived features, the perturbation magnitude is 
\begin{equation}
\epsilon=
\max\!\left[
0.1\sigma,\;
0.05\,(g^{\max}-g^{\min})
\right] \;,
\label{eq:epsilon}
\end{equation} 
where $\sigma$ is the standard deviation of feature $g$,
over the training set. Here $g^{\max}$ and $g^{\min}$
are the maximum and minimum observed values of $g$, respectively.
This choice yields a valid local threshold crossing, while remaining close to the decision boundary.

\begin{figure}[t]
\centering
\includegraphics[width=\columnwidth]{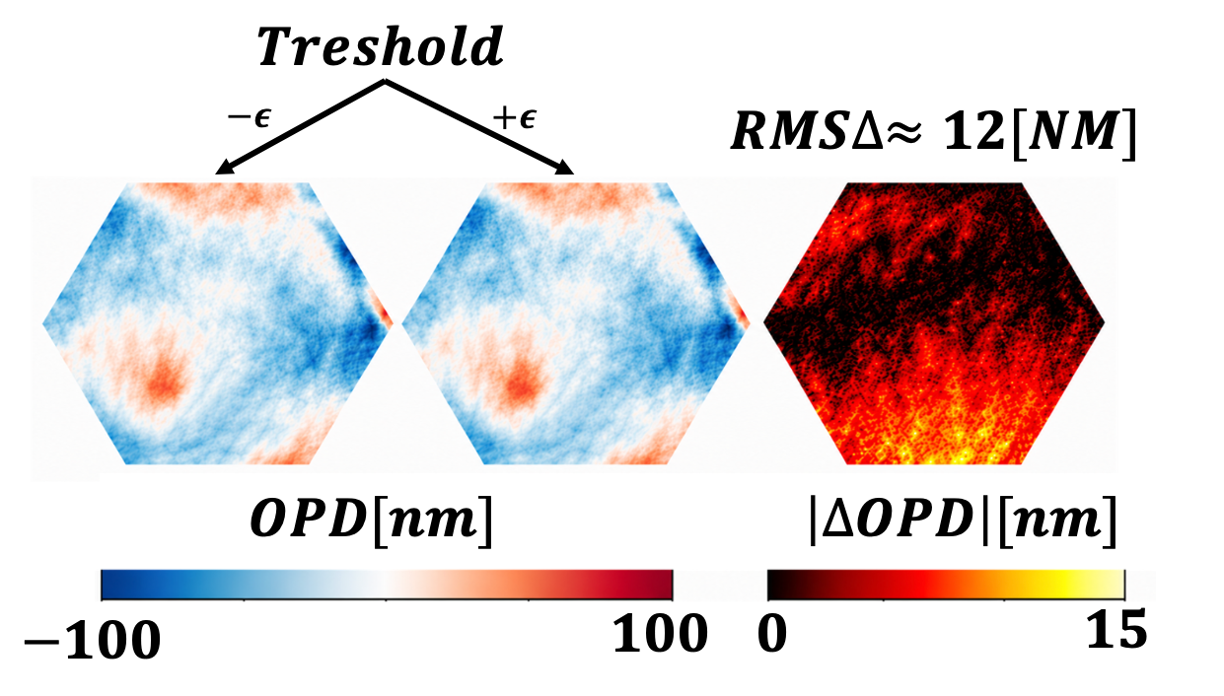}
\caption{
Threshold-crossing example for segment B4.
The left and center panels show inferred OPD maps obtained immediately below and above a threshold learned by GBR, respectively. The right panel shows the absolute difference map. The resulting discontinuity has
${\rm RMS}_{\Delta}(s)\approx 12$nm.
This example represents a relatively large perturbation response.
}
\label{fig:b4_threshold}
\end{figure}

Crossing a threshold at split $s$ yields an OPD discontinuity. Using
$g^{-}$ or $g^{+}$ yields, respectively, OPD maps denoted
$\widehat{\mathbf{OPD}}^{-}$ and
$\widehat{\mathbf{OPD}}^{+}$. Their difference is
\begin{equation}
\Delta_{\rm OPD}(s)
=
\widehat{\mathbf{OPD}}^{+}
-
\widehat{\mathbf{OPD}}^{-}
.
\label{eq:delta}
\end{equation}
The spatial RMS of $\Delta_{\rm OPD}(s)$ is denoted 
${\rm RMS}_{\Delta}(s)$. 
Figure~\ref{fig:b4_threshold} shows a representative example: 
segment~B4. We study here the dominant thermal-slope feature
\(r^{\mathrm{PDU}}_{\mathrm{slope},3h}\), corresponding to a telemetry
channel of the 3-hour temperature trend of the MIRI power-distribution-unit.
As shown in Fig.~\ref{fig:b4_threshold}, crossing the learned threshold
changes the reconstructed OPD only modestly. The two inferred OPD maps
differ by approximately ${\rm RMS}_{\Delta}(s)\approx 12$nm,
with $|\Delta_{\rm OPD}|$
remaining below \(15\,\mathrm{nm}\). This is well below the typical JWST
wavefront-maintenance scale of \(60\)-\(80\,\mathrm{nm}\).

\begin{figure}[t]
\centering
\includegraphics[width=\columnwidth]{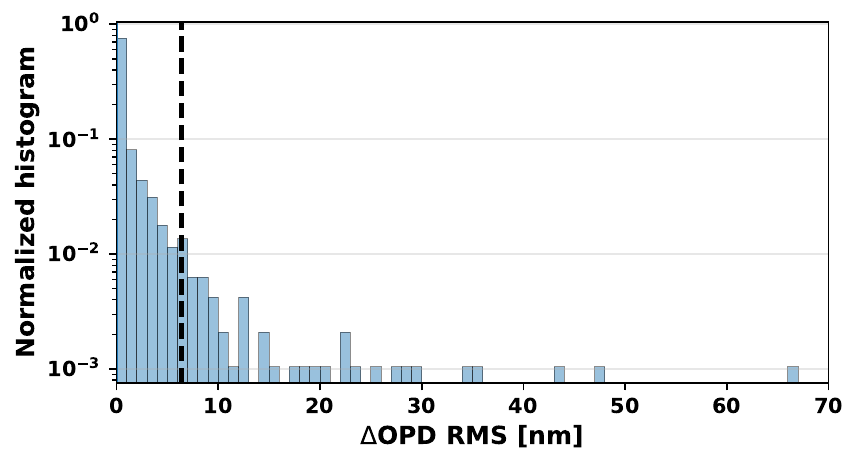}
\caption{
Normalized histogram of OPD discontinuities induced by threshold-crossing
perturbations. This histogram covers perturbation in all 18 mirror segments and all input features (telemetry and estimated piston).
The dashed line indicates the 95th percentile
(\(6.4\,\mathrm{nm}\)). % with 95\% of all discontinuities falling below it.
}
\label{fig:robustness_ecdf}
\end{figure}

More generally, we replicated this study across all mirror segments and features (telemetry and inferred pistons) of ${\bf g}$. 
Perturbations sometimes altered several internal tree decisions.
The resulting distribution of ${\rm RMS}_{\Delta}$ is shown in
Fig.~\ref{fig:robustness_ecdf}.
Across all perturbations, the median of  ${\rm RMS}_{\Delta}$ 
%$|\Delta_{\rm OPD}|$ 
 is
\(0.22\,\mathrm{nm}\). The 95th percentile is
\(6.4\,\mathrm{nm}\). 
Despite the threshold-based structure of individual trees,
the boosted ensemble exhibits gradual behavior under small
input perturbations: OPD jumps remained below typical JWST maintenance scales.  Our basic computer code and some data to run are appear in~\cite{github}.

% ======================================================================
%   6. DISCUSSION
% ======================================================================
\section{Discussion}
\label{sec:discussion}

\textbf{Toward predictive calibration scheduling.}\quad
Current WFSC scheduling relies primarily on fixed cadences and scalar RMS thresholds.
Telemetry driven OPD estimation suggests a complementary approach in which
predicted wavefront evolution is continuously monitored between dedicated
wavefront sensing observations. Calibration could then be advanced when drift
is accelerating and deferred when conditions remain stable. A simple decision
rule based on predicted full aperture RMS provides a practical mechanism for
anticipating threshold crossings rather than reacting to them after they occur.
At segments for which the model demonstrates statistical significance, telemetry-driven OPD estimation may provide an additional indicator of wavefront evolution between dedicated WFSC observations. Such information could help identify periods of elevated wavefront drift while reducing reliance on fixed calibration cadences. This suggests that routinely collected engineering telemetry may provide a useful low-cost signal for future operational calibration planning. 

\textbf{Thermal analysis from a learned model.}\quad
The features mostly relate to the thermal state and recent history of the telescope. 
The tree model points out features that relate more significantly to optical aberrations than others. This can provide insights into thermal analysis of the telescope, and engineering of future observatories. For some mirror segments, the existing telemetry may not provide statistical significance. So, future engineering may insert additional telemetry probes in some locations, to help inference of the OPD in these mirror segments.

\textbf{Limitations and generalizations.}\quad
In the limited data we have, segments with negative $\Eexpl$ seem to have a tendency  to cluster in specific time intervals. Potentially, there are unmodeled disturbances or operational conditions that are poorly represented in the training data.  The data we have does not isolate external disturbances that may contribute to OPD variability and are not represented in the available telemetry. 
Such disturbances may, potentially, relate to micrometeorites and space weather. 
Performance may also vary across observing campaigns and spacecraft operating modes, including momentum unloading events and guide star acquisitions. 
%Because both training and test labels are derived from the same WFSC measurement process and evaluated using a strict time-ordered split, calibration-specific pointing effects are treated consistently. Future work could explicitly model calibration attitude history and Sun-angle evolution.  

Longer temporal coverage may help to determine the extent to which the learned model remains valid across different operational regimes. Extending the analysis to multi year datasets would improve robustness and capture seasonal variability in the thermal state of the observatory.

To simplify the work, physical intuition lead us to some feature selections, specific $T_{\alpha}$ values and state statistics. Moreover, the model estimates OPD maps per time, rather than a dynamical system varying in time. 
Future work should be based on more data that will continue to be acquired during operation. This will enable a more systematic search across the full JWST housekeeping archive for good features. Increase in training data size will enable better models to train. In particular, exploring neural architectures is a welcome direction.

Future work may also extract features from data from auxiliary sensors that the astronomy community has, and that are not on JWST. These may include solar activity, space weather (charged particles) and clouds of inter-planetary dust. Extended data
may reveal stronger predictors, particularly for mirror segments that currently remain difficult to infer. Modeling may also generalize to explicitly model state dynamics.

The same principle may apply to other complex observatories, where
continuous knowledge of the wavefront state is a key operational requirement. These include other segmented telescopes and future large aperture space missions. 

% ======================================================================
%   7. CONCLUSION
% ======================================================================

% Acknowledgments only for camera-ready
\ifpeerreview \else
\section*{Acknowledgments}
This research is partly supported by the Israel Science Foundation (ISF grant 2514/23) and KLA. The work was
conducted in the Ollendorff Minerva Center. Minerva is
funded through the BMBF. 
Y.~Y.~Schechner is the Mark and Diane
Seiden Chair in Science at the Technion. He is a Landau
Fellow, supported by the Taub Foundation.  J. J. Wang acknowledges support from the Alfred P. Sloan Foundation. This work is based on observations made with the NASA/ESA/CSA James Webb Space Telescope. The data were obtained from the Mikulski Archive for Space Telescopes at the Space Telescope Science Institute, which is operated by the Association of Universities for Research in Astronomy, Inc., under NASA contract NAS 5-03127 for JWST. These observations are associated with program 7942. Support for program 7942 was provided by NASA through a grant from the Space Telescope Science Institute, which is operated by the Association of Universities for Research in Astronomy, Inc., under NASA contract NAS 5-03127. A. Levis's work is supported by The Natural Sciences and Engineering Research Council of Canada (NSERC).
\fi

\bibliographystyle{IEEEtran}
\bibliography{references}

\end{document}